\documentclass{article}

 \usepackage[dblblindworkshop, final]{neurips_2025}
\workshoptitle{SPIGM}

\usepackage[utf8]{inputenc} 
\usepackage[T1]{fontenc}    
\usepackage{hyperref}       
\usepackage{url}            
\usepackage{booktabs}       
\usepackage{amsfonts}       
\usepackage{nicefrac}       
\usepackage{microtype}      
\usepackage{xcolor}         
\usepackage{amsmath}
 \usepackage{graphicx}
 \usepackage{subcaption}
 \usepackage{tabularx}
\usepackage{booktabs} 
\usepackage{caption}
\usepackage{array}
\usepackage{dashrule}
\usepackage{arydshln} 
\usepackage{calc} 
\usepackage{wrapfig}

\def\approxprop{%
  \def\p{%
    \setbox0=\vbox{\hbox{$\propto$}}%
    \ht0=0.6ex \box0 }%
  \def\s{%
    \vbox{\hbox{$\sim$}}%
  }%
  \mathrel{\raisebox{0.7ex}{%
      \mbox{$\underset{\s}{\p}$}%
    }}%
}
\title{Bayes-PD: Exploring a Sequence to Binding Bayesian Neural Network model trained on Phage Display data}

\author{
  Ilann Amiaud--Plachy \\
  InstaDeep\\
  Paris, France \\
  \texttt{i.amiaud-plachy@instadeep.com} \\
  \And
  Michael Blank \\
  BioNTech \\
  Munich, Germany \\
  \texttt{michael.blank@biontech.de}  \\
  \And
  Oliver Bent \\
  InstaDeep \\
  Paris, France \\
  \texttt{o.bent@instadeep.com}  \\
  \And
  Sebastien Boyer \\
  InstaDeep \\
  Paris, France \\
  \texttt{s.boyer@instadeep.com} \\
}

\begin{document}

\maketitle

\begin{abstract}

  Phage display is a powerful laboratory technique used to study the interactions between proteins and other molecules, whether other proteins, peptides, DNA or RNA. The under-utilisation of this data in conjunction with deep learning models for protein design may be attributed to; high experimental noise levels; the complex nature of data pre-processing; and difficulty interpreting these experimental results. In this work, we propose a novel approach utilising a Bayesian Neural Network within a training loop, in order to simulate the phage display experiment and its associated noise. Our goal is to investigate how understanding the experimental noise and model uncertainty can enable the reliable application of such models to reliably interpret phage display experiments. We validate our approach using actual binding affinity measurements instead of relying solely on proxy values derived from `held-out' phage display rounds.
\end{abstract}

\section{Introduction}

\begin{figure}
    \centering
    \includegraphics[width=0.65\linewidth]{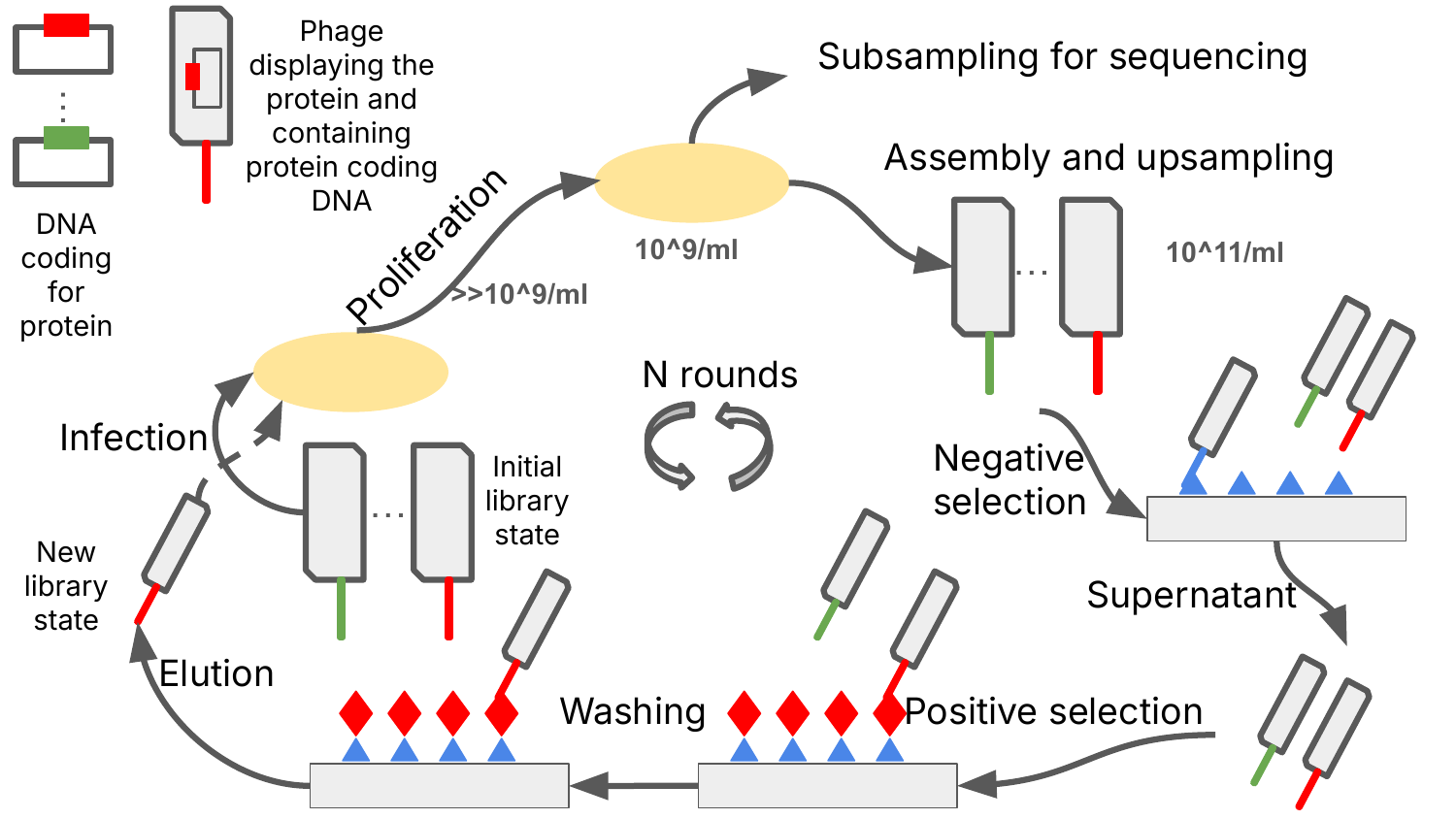}
    \caption{\textbf{Schematic of a selection step in phage display.}}
    \label{fig:Phage_Display}
\end{figure}

Phage display is a high-throughput experimental technique used to screen large protein libraries for their ability to bind to a specific target \citep{SONG2024128455}\citep{Grzegorz2021}. These libraries typically consist of millions of different proteins, with each protein being present in millions of copies at the start of the experiment. The phage display experiment provides a proxy measure of binding known as selectivity, which represents the change in sequence abundances (or frequencies) before and after the selection process \citep{hierarchy_and_extremes}.

\begin{equation}
\label{eq:select}
    s_i^{N}=\frac{f_i^{N+1}}{f_i^{N}} \approxprop \mathrm{binding_{affinity}}
\end{equation}

where $i$ refers to sequence $i$, $N$ to the selection step, $f_i^N$ is the frequency of sequence $i$ in the total population at selection step $N$ and $\approxprop$ refers to approximately correlated as the usual phage display selection step contains more than the selection for the designated target (see negative selection in Figure \ref{fig:Phage_Display}).\\

Although the results of a phage display experiment consist of pairs of integers representing the counts of sequences before and after selection (obtained through high-throughput sequencing, an experimental set up allowing for the reading of hundreds of millions of sequences at once \citep{Zeng2020}), these numbers need to be transformed into frequency comparisons. This transformation is necessary for two main reasons: first, the initial counts (before the selection step) are not uniformly distributed, and second, multiple sampling steps from those libraries occur before sequencing, making the absolute values of these integers less informative.\\

Due to the inherent randomness and noise associated with the binding process (see Appendix Figure~\ref{fig:noise}), along with counting noise and multiple sampling steps involved in the experiment, it is essential to develop a model that can accommodate these intricacies in both its architecture and training. This need has only recently begun to garner interest \citep{rivoire_phage_display}.
Even in recent undeniably successful research contributions, the outputs from phage display have been in one case, only utilized to train a binder-non-binder classifier using only the results from the final selection step \citep{Gfeller_Science_2025}, while an other approach has attempted to directly regress on selectivity \citep{Ito31122023}. However, we argue that this method of selectivity regression overlooks the actual structure of the experimental setup, which relies on integers. Consequently, it likely fails to account for counting noise.
In parallel, the accessibility of deep Bayesian modelling has been increasing, largely due to the availability of Python libraries such as PyMC3 \citep{salvatier2015probabilisticprogrammingpythonusing} and Pyro \citep{bingham2018pyro}. These libraries provide a framework for training deep Bayesian models. 
Additionally, they offer ready-to-use training strategies. For example, Pyro includes scalable Variational Inference (SVI) and simple yet effective variational distributions, like the multidimensional Gaussian with a diagonal covariance matrix. This functionality enables exploration of models at a scale at which deep learning could be considered. Indeed when using a diagonal Gaussian variational distribution, models with millions of parameters can be trained. 
\\
While implementing and training Bayesian deep learning models has become easier, progress on the front of their explain-ability has also seen great advances \citep{bykov2021explainingbayesianneuralnetworks}.\\
Here we explore how by training a Bayesian deep neural network sequence to binding probability model within a dedicated training loop simulating the phage display selection experiment, we could leverage both our understanding of the model uncertainty and model output to propose, with high confidence, sequences within a known range of binding affinity.\\
Additionally, our model incorporates strategies for scalability regarding speed and memory management, particularly in cases where effective diversity during selection rounds presents challenges.\\
Finally, since our model is validated using actual binding affinity measurements instead of selectivities from phage display experiments, we gain valuable insights into the limitations of both the modelling and the experiments. We have addressed these shortcomings through a series of possible model enhancements and how we could make them work.
\section{Methods}
\subsection{Datasets}
\label{sec:dataset}
We have access to 3 different phage display experiments performing selection on 3 different targets. Hence we will have $\mathrm{experiment_c}$ related to selection on $\mathrm{target_c}$,~$c$ going from 1 to 3. Schematics of the different experiments are presented in appendix Figure \ref{fig:experiment1}-\ref{fig:experiment3}.\\
All the phage display data used are round 2 and 3 of the selection process: we assume that the rounds of selection from 1 to 2 are too noisy and will hardly reflect changes in frequency useful for computing binding affinity proxies. Indeed, given low initial counts at round 1 and stringent selection/sampling leading to round 2, a lot of those changes would be mostly accidental or an example of poorly estimated selection.\\
We only have access to the change in frequency due to the overall selection process which is a convolution between a negative selection step (to ensure that proteins are not selected because they bind to the bead-selection matrix), and a positive selection step for the target. We do not have access to data enabling a deconvolution of these two steps.\\
We have access to 3 "replicates" of selection ($\mathrm{experiment_1}$: Exp 1: round 2 --> round 3a, Exp 2: round 2 --> round 3b and Exp 3: round 2 --> round 3c). Those are not real replicates as they differ from each others by the concentration of target being used, yet we show in Appendix (Figure~\ref{fig:selectivity}) that they could loosely be used as such since those experiments output similar selectivities. For $\mathrm{experiment_1}$ we additionally have access to technical replicates (Exp 4-6) from resequencing of the rounds. All those replicates share the same round 1 and 2. We always use the 2 highest concentrations for training (round 3a and round 3b) and the lowest one (round 3c) for validation. This makes the validation and the training set quite correlated and so we will not place too much incentive on the difference in metrics between these 2 splits. Although it is still useful to look at them through a qualitative lens. Finally, by choosing the lowest target concentration as our validation, we hope to stay away as far as possible from the training set and put the model in the hardest validation mode.\\
For $\mathrm{experiment_1}$ we also have access to a test set made of 126 sequences which have been selected via bioinformatic NGS-data analysis from $\mathrm{experiment_1}$ as well as sequences redesigned by a Bayesian Flow  Network (BFN) \citep{Atkinson2025}, fine tuned on the output of the phage display (see appendix \ref{section:BFN}), and for which actual binding affinity measurements have been performed. We will consider this set as the appropriate way to test our models.

\subsection{Model}


A key strength of Bayesian modelling, which extends beyond the neural network architecture itself, is the ability to model stochastic processes more precisely. This detailed representation is then directly incorporated into the calculation of posterior distributions. A visual representation of our model is provided in Figure~\ref{fig:representation}, while its Bayesian representation is available in Appendix Figure~\ref{fig:bayes}.

\subsubsection{Sequence Pre-processing}


Our dataset consists of raw protein sequences that require encoding. To this end, we used a sequence embedder built from a protein language model. After evaluating several state-of-the-art protein LLMs on their metric performance, inference time and memory utilization, we selected the lightest ESM-2 transformer (8 million parameters) \citep{Lin2022.07.20.500902}. This model, which embeds each amino acid in a 320-dimensional vector, was chosen for its effectiveness compared to larger models, facilitating fast and memory-efficient training. 

\subsubsection{Faithful Modelling of Phage Display Experiments}

To closely replicate the biological experiment, our model is designed to take two inputs: the protein sequences and their corresponding counts at step N. The model's output is a predicted count for each sequence after the selection process. This prediction is then compared against the ground truth count observed at step N+1.\\
A key challenge is that the observed sequence counts are several orders of magnitude smaller than the total biological population. To faithfully model the uncertainty associated with this subsampling, our workflow involves three steps. First, we upsample the input counts from step N to the estimated total population size. Second, we apply our selection model at this population scale. Finally, we downsample the predicted post-selection counts to the sequencing scale to generate the final output.\\
Furthermore, to accurately represent the stochasticity of the subsampling process, we incorporate a probabilistic sampling step into our model. The multinomial distribution is a natural choice for this task, as it can model the selection of counts based on the predicted relative abundance of each sequence.\\
Initial models were built using this multinomial distribution and were successfully trained on small-scale experiments. However, the multinomial approach presents a significant computational challenge: the inherent dependency between sequence counts requires all sequences to be processed simultaneously. This is computationally prohibitive for datasets with a large number of unique sequences, leading to memory issues within the neural network. To overcome this limitation, we demonstrate that the multinomial distribution can be effectively approximated by a set of independent Poisson distributions. This approximation, known as the law of rare events, holds in our context of many sequences with low individual probabilities. Adopting the Poisson approximation provides multiple advantages: it enables mini-batch training, simplifies the model mathematically, and obviates the need to compute relative abundances. Instead, the predicted count for each sequence directly serves as the rate parameter $\lambda_{i}^{N}$ for its respective Poisson distribution (see Appendix~\ref{sec:mult_to_poisson}).

However, even with the Poisson distribution, implementing batch-based training introduces an approximation. The normalization of the model's output to the scale of the sequenced N+1 population requires the total sum of predicted counts across the entire dataset. When using mini-batches, this global sum must be estimated from the counts within the current batch. Consequently, the batch size cannot be excessively small; a sufficiently large batch is necessary to ensure this estimation is accurate and to maintain a training consistency comparable to the multinomial approach. The combination of Poisson's law and reasonable batch size not only allows for better results than the multinomial case but also regularizes the model by leveraging the estimated total population size confidence. A study on batch size is provided in Appendix~\ref{sec:batch}.

\begin{figure}
    \centering
    \includegraphics[width=0.85\linewidth]{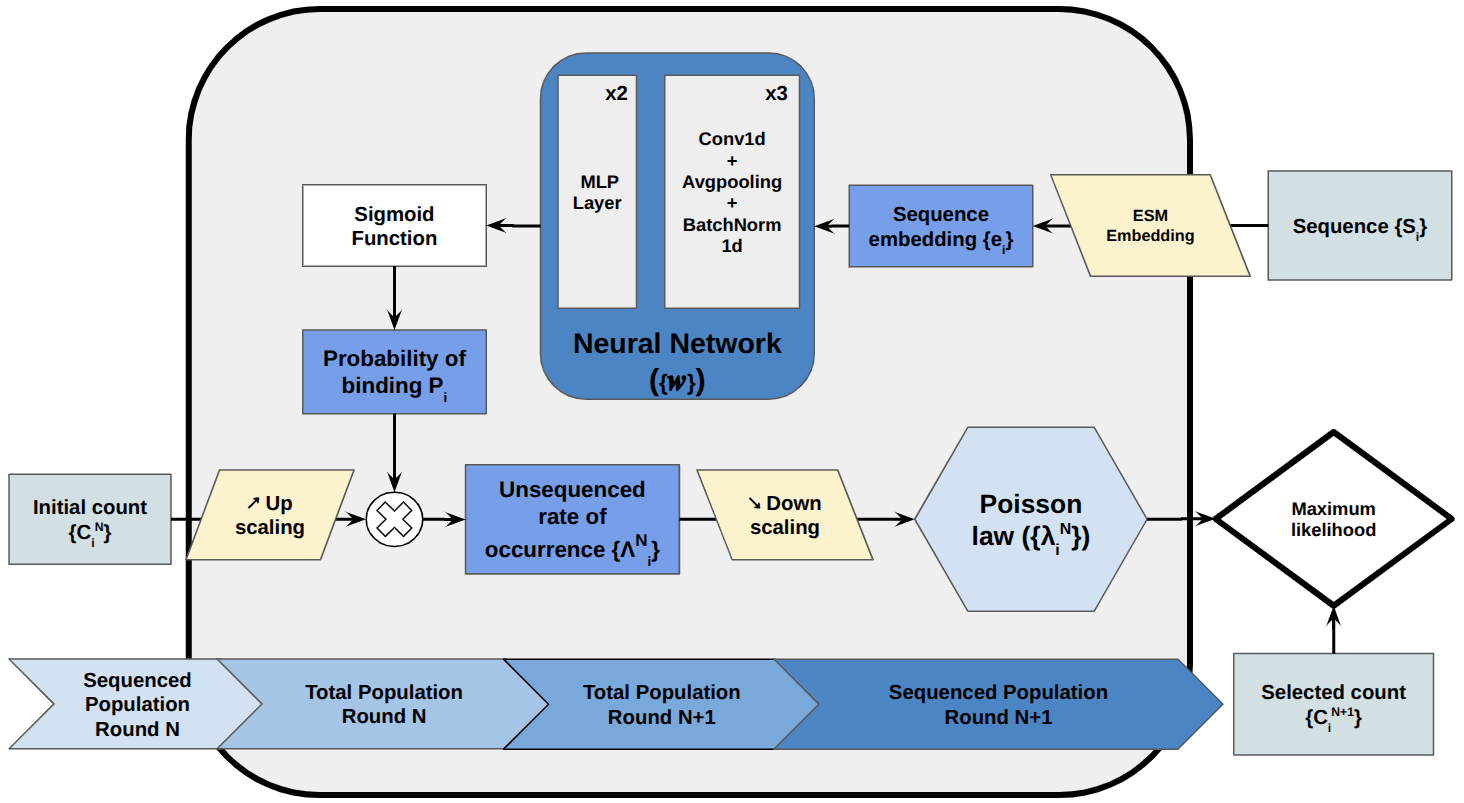}
    \caption{\textbf{Phage display Poisson model}. Unsequenced rate of occurrence $\Lambda^N_i$ stands for the parameters of our Poisson law at the relevant experimental population size, before downscaling to obtain the sequenced rate of occurrence $\lambda^N_i$ which is measured (sequenced).
    }
    \label{fig:representation}
\end{figure}

\subsubsection{Bayesian Neural Network}


Our model is designed to infer the binding probability to the specified target for each sequence. These inferred probabilities, when combined with the upsampled initial counts, provide the rate parameters for the corresponding Poisson distributions.\\
Given the sequential nature and contextual information inherent in the sequence embeddings, a Convolutional Neural Network (CNN) was selected as the core architecture. The training of Bayesian neural networks can be unstable; therefore, balancing the number of model parameters is crucial to prevent training collapse. CNNs provide an effective balance in this regard. In contrast, alternative architectures like Multi-Layer Perceptrons (MLPs) were deemed less suitable, as they either perform poorly with few parameters or fail to converge when the network is too wide or deep.\\
Our specific architecture, consisting of three convolutional layers with batch normalization and average pooling, is based on the work of \citep{Chandra2023PepCNN}. To further improve training stability and activate data reconstruction, we implemented standard variational optimization techniques, including loss scheduling and Kullback-Leibler (KL) annealing \citep{fu2019cyclicalannealingschedulesimple}. Finally, the hyper-parameters were tuned on the smallest dataset to establish a robust baseline model that demonstrates strong performance across all datasets. Depending on the underlying correlation within the dataset, some hyperparameters choice can be crucial, such as changing the activation function: for instance, ReLU activation tends to be robust, and tanh will be more sharp to activate the learning. More details about the training process can be find in Appendix~\ref{sec:training}.

\section{Results}

Quantifying the model's performance is complex due to the highly noisy nature of the dataset, the fact that raw counts do not directly convey the underlying biological insights, and strong correlation between our splits. In this light, we decided to use, as a validation of our method, a Test dataset of experimentally derived binding affinity. This dataset, hence, shares little in term of noise and experimental set up with the phage display dataset, except that its rigorous way of measuring binding affinity should be related to the selection process at play in the phage display experiment. 



\subsection{Results Table}

The performance of our baseline model, with fixed hyper parameters, is reported in Table~\ref{tab:corr_results}.

\begin{table}[h!]
    \centering
    \begin{tabularx}{\textwidth}{|l|*{8}{>{\centering\arraybackslash}X|}}
        \hline
        \multicolumn{1}{|c|}{} & \multicolumn{3}{c|}{\textbf{Train}} & \multicolumn{3}{c|}{\textbf{Valid}} & \multicolumn{2}{c|}{\textbf{Test}} \\
        \cline{2-9}
        \multicolumn{1}{|c|}{\textbf{Targets}} & $C_i^{\text{N}}$ vs $C_i^{\text{N+1}}$ & $C_i^{\text{N+1}}$ vs $C_i^{\text{Pred}}$ & $P_i$ vs $s_i$ & $C_i^{\text{N}}$ vs $C_i^{\text{N+1}}$ & $C_i^{\text{N+1}}$ vs $C_i^{\text{Pred}}$ & $P_i$ vs $s_i$ & $K_{d,i}$ vs $s_i$ & $K_{d,i}$ vs $P_i$ \\
        \hline
        \textbf{Target 1} & 0.41 & 0.62 & 0.50 & 0.41 & 0.57 & 0.42 & -0.24 & -0.35 \\
        \hline
        \textbf{Target 2} & 0.23 & 0.44 & 0.31 & 0.24 & 0.37 & 0.22 & * & * \\
        \hline
        \textbf{Target 3} & 0.20 & 0.43 & 0.46 & 0.19 & 0.47 & 0.45 & * & * \\
        \hline
    \end{tabularx}
    \vspace{1em}
    \caption{\textbf{Spearman correlation metrics for the baseline model across different datasets and targets}. The Test set evaluates generalization to a distinct but closer to ground truth experimental set up, that is only available for target 1, as mentioned in Subsection~\ref{sec:dataset}.
    It is worth noting that the \( s_i \) set contains the natural sequences, which are also included in the \( K_d \) set, while $K_d$ set further includes generated sequences.
    So, the gain in performance can partially be explained by the experimental characterization of those generated sequences. For each target, the model was trained on a distinct data partition. (See Appendix~\ref{app:experimental} for experimental setup.)} 
    \label{tab:corr_results}
\end{table}

For the \textbf{Train} and \textbf{Validation} sets, each are reporting only one experiment from their set to avoid pooling the results, and we report two key performance metrics. The first is the correlation between the model's predicted binding probabilities ($P_i$) and the experimentally derived selectivities ($s_i$). As selectivities represent a meaningful global statistic, this metric assesses how well the model captures underlying data properties. The second metric is the correlation between the predicted counts ($C_i^{\text{Pred}}$) and the ground truth counts ($C_i^{\text{N+1}}$). This directly evaluates performance on the primary data reconstruction task and allows for a clear comparison against the \textbf{null model} to quantify the benefits of our learning approach. Null model is the correlation between the counts at round N and N+1 ($C_i^{\text{N}}$ vs $C_i^{\text{N+1}}$), giving us insight on the strength of the selection process in the dataset, because it underlines the change in repartition count during the round.\\
The \textbf{Test} set is used for a critical biological validation. Here, we evaluate the correlation between the dissociation constant ($K_{d,i}$) and the predicted binding probability ($P_i$). A strong \textbf{negative correlation} is biologically expected, as a higher dissociation constant (lower binding affinity) should correspond to a lower binding probability. Therefore, a more negative correlation coefficient indicates a more biologically sound and meaningful model. All correlations are calculated using \textbf{Spearman's rank correlation coefficient} to robustly handle the non-linear relationships inherent in the data.\\
Our model's reconstruction performance is significantly better than that of the null model (Table ~\ref{tab:corr_results}): demonstrating that our model learns beyond the simple correlation of frequencies between rounds of selection.\\
Our correlation with the test set represents an improvement from what is directly accessible from the data (using selectivities, Table ~\ref{tab:corr_results}). Comparing only to sequences from the phage display experiment (Figure~\ref{fig:scatter_plot}) we still see an improvement in our correlation (from -0.24 to -0.33) compared to using selectivities. When looking only at generated sequences this correlation reaches -0.46, showcasing the model ability to generalize to unseen sequences (hamming distances from seed sequences varies between 1 to 10).

\subsection{Scatter plots}

\begin{figure}
    \centering
    \includegraphics[width=0.95\linewidth]{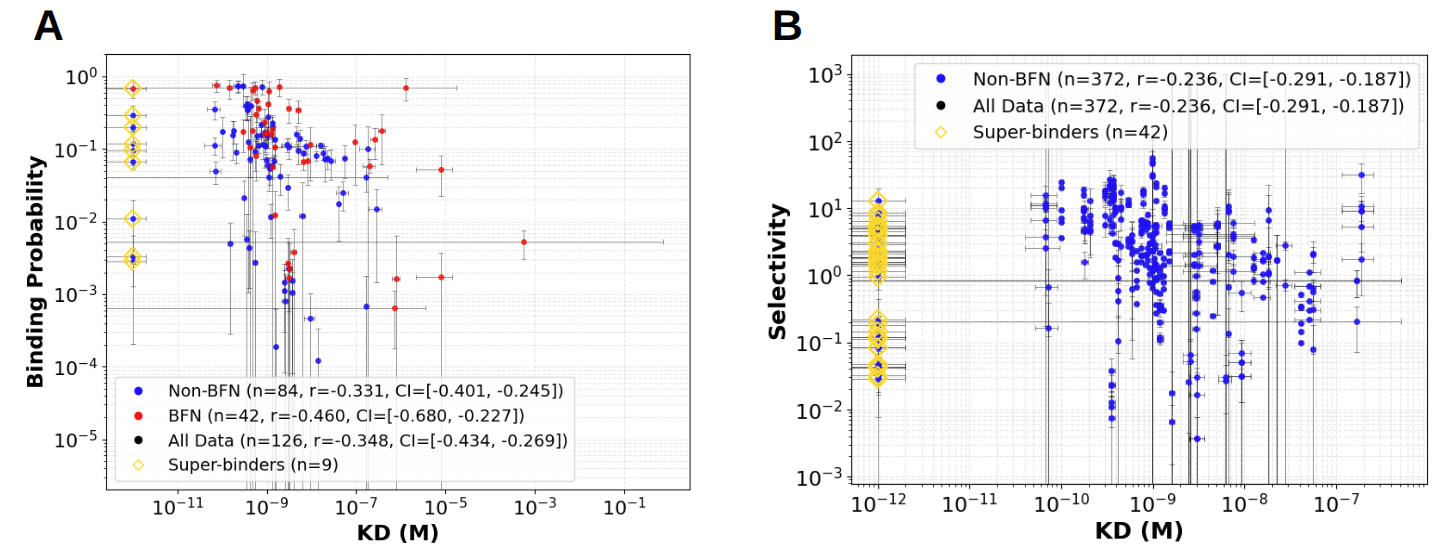}
    \caption{ \textbf{Correlation plots with the dissociation constant test set} (\textbf{A}) \textbf{ for} $\mathbf{target_1}$: Correlation with model prediction. Error bars on predicting binding probability (y axis) are estimated errors from N samples of the models while the actual dot markers represent the estimated mean from that same sampling. Error bars on $K_{d}$  are uncertainty from curve fitting. (\textbf{B}) : Correlation with raw Selectivities (Exp 1-6). Error bars on selectivity are estimated from counting noise following $\frac{\Delta s}{s} = \frac{1}{\sqrt{C_i^N}} +\frac{1}{\sqrt{C_i^{N+1}}}+\mathcal{O}(C_{tot}^N)$. Error bars on $K_{d}$ are uncertainty from curve fitting. 95\% Confidence intervals on correlation are based on 97.5\% and 2.5\% percentiles of N samples of the model compared to N Gaussian samples of the $K_{d}$ values. (In our case, N = 1000).}
    \label{fig:scatter_plot}
\end{figure}

Analysis of the scatter plots depicted in Figure~\ref{fig:scatter_plot} reveals key insights into the model's behaviour, particularly regarding predicted binding probabilities and experimental selectivities. The plot correlating these two metrics is densely populated, as it aggregates data from multiple experiments 
.\\
The model struggles to correctly predict the behaviour of ``super-binders'' --- sequences with extremely high affinity, capped by an instrument measurement floor of $10^{-12}$.
Two main hypothesis could explain this behaviour. The first and possibly more likely is a caveats of our $K_d$ measurement which could via dimerization of the VHHs inflate their low $K_d$. It is thus possible that those super binders are, in fact, prone to such interaction. This is not testable within our model as it does not relate to phage display. The other possibility is that those sequences are also exhibiting a strong affinity for the non-target bead-selection matrix, leading to their elimination during negative selection. This creates a conflicting signal: the strong negative signal can effectively cancel out the positive signal, resulting in an erroneously low predicted binding probability.\\
Despite this specific limitation, the overall correlation plots indicate that the model's output is well-structured and successfully captures specific sequence-target binding events.

\subsection{Explainability (XAI)}

A primary objective of this study was to identify the most influential amino acid sites for binding to a specific target. To achieve this, we applied the method developed by \citep{bykov2021explainingbayesianneuralnetworks} to our Bayesian model, which leverages the posterior distribution to generate robust feature attributions. This approach involves sampling multiple deterministic networks from the learned posterior, generating an explanation for each network using an XAI method such as Integrated Gradients (see \citep{pmlr-v70-sundararajan17a}), and then aggregating these individual explanations.

\begin{figure}
    \centering
    \includegraphics[width=0.8\linewidth]{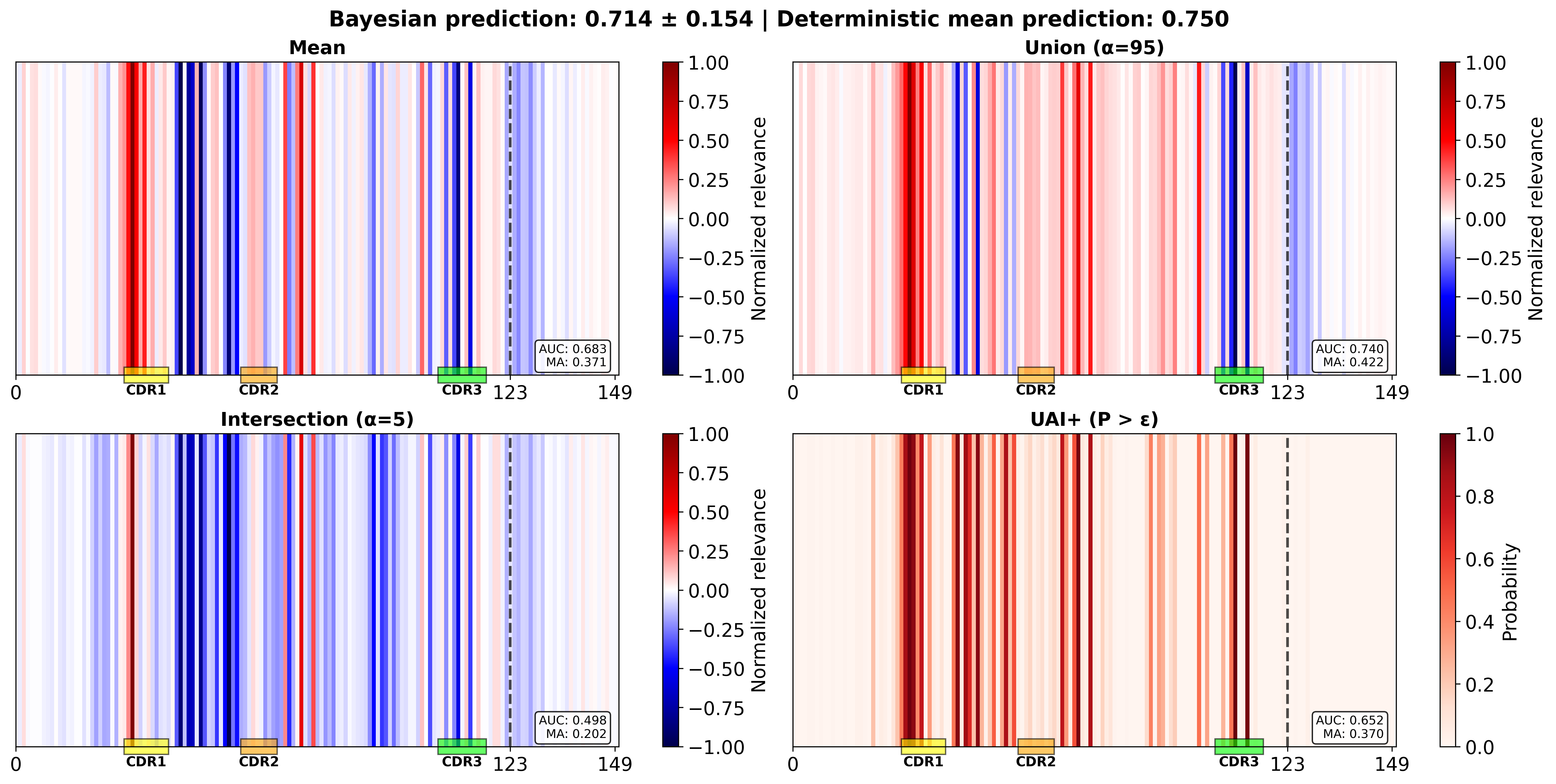}
    \caption{\textbf{Integrated Gradient bayesian explanation of a sequence}. Dash line is a visual representation of the end of the sequence: sequences are batched at prediction time and thus need padding.}
    \label{fig:explanation}
\end{figure}

An example of such an explanation is shown in Figure \ref{fig:explanation}. These visualizations highlight the specific residues that the model utilized for its predictions. The figure also overlays the Complementarity-Determining Regions (CDRs) \citep{Wong2019}, which are theoretically the primary sites of interaction for antibodies which are our proteins of interest, especially true for CDR3. The visualization shows that while the explanatory signal is not confined exclusively to the CDRs, they constitute a significant portion of the attribution. Building upon the methodology proposed by \citep{bykov2021explainingbayesianneuralnetworks}, we employ a quantitative evaluation framework to assess our explanations. We compute the Area Under the Receiver Operating Characteristic (AUC-ROC) curve by sweeping an absolute relevance threshold to gauge the overall quality of the attributions. Additionally, we calculate the Relevance Mass Accuracy, a metric that quantifies the portion of total relevance concentrated within the Complementarity Determining Regions (CDRs). By averaging these attribution scores across all sequences, we can identify the sites that are globally most critical for the binding phenomenon.
 
\subsection{Error Handling}

A second objective was to predict the binding affinity of novel sequences. While the model's predictive performance was varied, our Bayesian framework provides a distinct advantage: the ability to quantify the uncertainty associated with each prediction. We estimate the mean and standard deviation of any prediction by sampling multiple times from the model's posterior distribution. The relative standard deviation serves as a normalized uncertainty metric for each prediction.\\
To demonstrate the relationship between model uncertainty and accuracy, we stratified the test set into three groups (terciles) based on this uncertainty metric. The limited size of the test set prevented stratification into more groups. A similar analysis was performed using an AUC ROC metric, which evaluates how well the model's explanations distinguish CDR sites from non-CDR sites. Stratifying this metric by uncertainty provides insight into whether the model is more confident when it correctly focuses on the CDRs. The results of these stratified analyses are presented in Figure \ref{fig:correlation-bars}. (See Appendix Figure \ref{fig:auc_comparison} for the other Explainability metrics).\\
In conclusion, this study demonstrates that by combining quantified uncertainty with model explanations, we can effectively assess the reliability of individual predictions (Top k enrichment study are available in Appendix Figure \ref{fig:top_k_enrichment}, \ref{fig:top_k_enrichment_auc}). Although the confidence interval length remains challenging due to high intrinsic noise, our approach allows us to identify a subset of high-confidence results. The convergence of low uncertainty scores with biologically plausible explanations provides a strong filter for the most trustworthy predictions, offering a practical strategy for extracting reliable insights from complex and noisy data.

\begin{figure}
    \centering
    \includegraphics[width=0.9\linewidth]{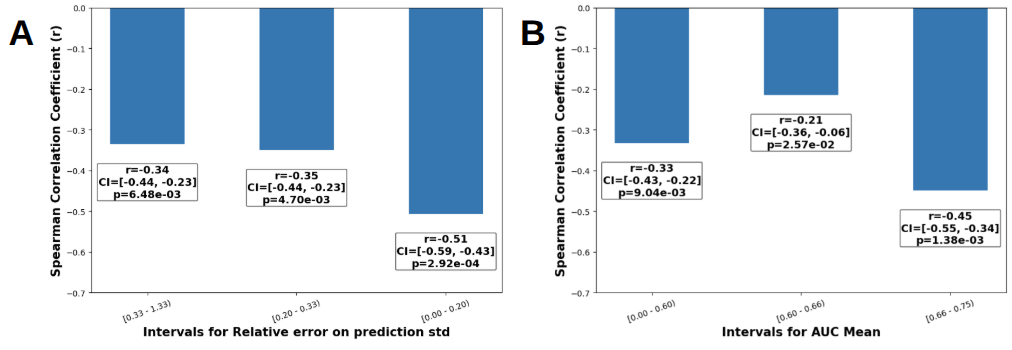}
    \caption{\textbf{Error handling possibilities within the Bayesian model.}  
    (\textbf{A}): Error handling based on relative standard deviation. (\textbf{B}): Error handling based on AUC Mean metric}
    \label{fig:correlation-bars}
\end{figure}

\section{Discussion}
\label{Sec:Discussion}
In most cases, phage display experiments are not designed to serve as a training dataset for deep learning-based modelling, even though they could be used for that purpose in principle given some slight modifications. Typically, these experiments are intended to identify interesting clones. Given this objective, it makes sense to adjust the concentration of the target between selection rounds. The same reasoning applies to the omission of recording negative selections.\\
We believe that the last point is crucial for correctly interpreting the phage display data, especially if the goal is to train a deep learning model. This understanding might help us identify the very strong binders that our model overlooked (indicated by the yellow squares in Figure~\ref{fig:scatter_plot}). First, $K_d$ data points only account for target binding, but our binding probability accounts for both negative and positive selection, with negative selection accounting for base binding, and positive selection to both base and target binding. We attempted to incorporate negative selection as an additional latent variable alongside positive selection to explain the outcomes of the phage display experiment. However, we were unable to break the symmetry between these two latent variables ($P^{\text{neg}}_i$ and $P^{\text{pos}}_i$) during training:

\begin{equation}
\label{eq:pos}
    P_i = P^{\text{pos}}_i \times (1 - P^{\text{neg}}_i)
\end{equation}

The results showed that our model would focus either on the positive or the negative term, and wouldn't separate the two phenomenons (Appendix~\ref{section:multiple}).

We theoretically broke the symmetry of the selection process by using a model based on the Boltzmann law, a methodology inspired by \citep{rivoire_phage_display}. Our architecture uses a flexible number of networks (1, 2, or 4) to generate four latent variables that represent distinct binding modes. This has a lot of similarities with a softmax output, that should be well fitted for our neural network. Specifically, the terms $e^{-E_x, i}$ and $e^{-E_{nx, i}}$ describe the binding and non-binding modes for both positive and negative selection, respectively, with $x$ being $b$ for the base and $t$ for the target.\\
The calculation of probabilities is defined by two key equations:
\begin{itemize}
    \item \textbf{Negative Selection}: This phase only considers the binding to the base, as shown in the probability equation below.
    \item \textbf{Positive Selection}: This phase includes both the base and our target, and the probability is thus calculated by integrating signals from all four possible modes.
\end{itemize}

\begin{align}
P_{i}^{\text{neg}} &= \frac{e^{-E_{b, i}}}{e^{-E_{b, i}} + e^{-E_{nb, i}}}, &
P_{i}^{\text{pos}} &= \frac{e^{-E_{b, i}} + e^{-E_{t, i}}}{e^{-E_{b, i}} + e^{-E_{nb, i}} + e^{-E_{t, i}} + e^{-E_{nt, i}}}
\end{align}

Increasing the number of latent variables rapidly expands the solution space, rendering it intractable and prone to finding non-biological solutions. 

Despite this added complexity, the model persistently fails to separate the positive and negative selection signals, which we observe are deeply and intrinsically intertwined within the experimental data, as shown in the Appendix~\ref{section:multiple}. Therefore, we believe that sequencing the data right after negative selection, or having more rounds to reduce the impact of negative selection may be the only viable approach moving forward (see \citep{rivoire_phage_display}).
Finally, when combining different rounds of training or pooling together entirely different experiments while conditioning our model on the target, it is crucial to ensure that the predicted probabilities share the same scale. One way to achieve this is by incorporating a measurement of the final population size at the end of the selection phase, by normalizing by the sum of our model's output.

\begin{equation}
\forall a \in \mathbb{R}, \quad C_{i}^{N+1} = C_{\text{tot},i}^{N+1} \cdot \frac{a \cdot C_{\text{out},i}^{N+1}}{\sum_{j} a \cdot C_{\text{out},j}^{N+1}}
\end{equation}

Without this constraint and in this context, since the probabilities are learned as a near multiplicative constant in order to form a frequency, the solution state is too wide and the model's learning could significantly suffer.
\section{Conclusion}
We have trained a Bayesian Deep learning sequence to binding affinity scorer, carefully using Phage Display data.
Taking advantage of the probabilistic nature of our model as well as its interpretation, we identified ways to maximize our chance to pick sequences with reliable binding estimates matching actual binding affinity measurements.
In its actual state, our model and strategy might miss a number of good binders, but would rarely lead to the selection of sequences with bad binding properties.
We also discussed and explored solutions around mismatches between how Phage Display data is usually produced and the optimal way they could be generated for model training.
\bibliographystyle{plain}
\bibliography{references}

\clearpage

\appendix

\section{Training}
\label{sec:training}

\begin{figure}
    \centering
    \includegraphics[width=0.75\linewidth]{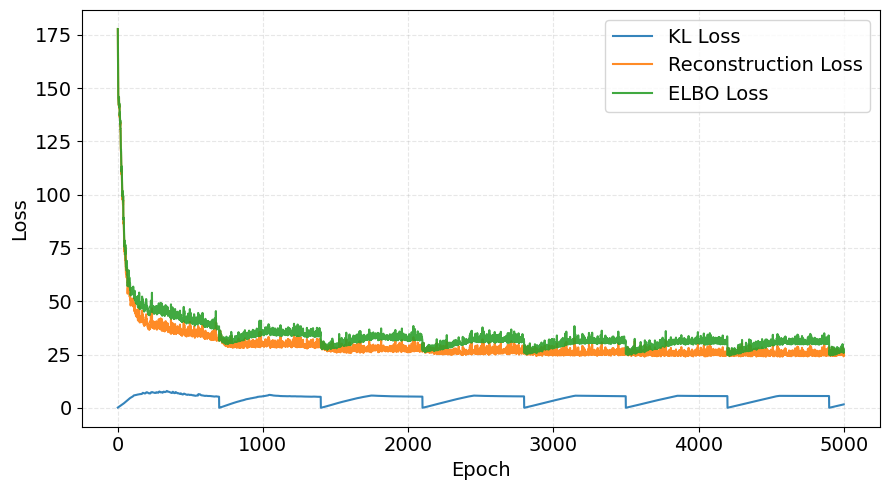}
    \caption{\textbf{Loss tracking during training, with ELBO Loss decomposition}}
    \label{fig:training_phase_losses}
\end{figure}

\begin{figure}
    \centering
    \includegraphics[width=0.75\linewidth]{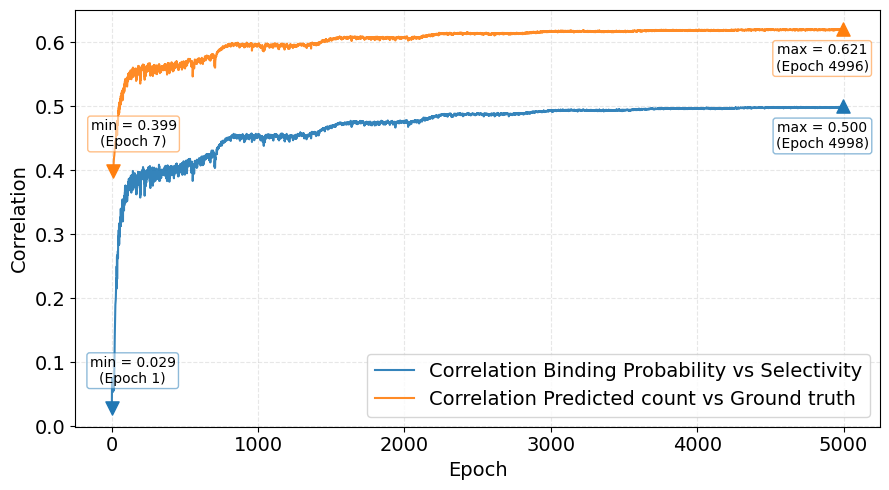}
    \caption{\textbf{Training Correlation metrics during training}}
    \label{fig:train_corr}
\end{figure}

The Bayesian model was trained using Stochastic Variational Inference (SVI), a scalable method for approximating the intractable posterior distribution of the network's weights. For this process, we employed a multivariate Normal variational distribution as our guide, which means that each of our weight distributions follows independent Normal laws. While we also experimented with more complex guides that account for dependencies between weights, they did not yield better results and came with a significant increase in computational cost.

To enhance performance and stability, the CNN architecture includes common layers such as a ReLU activation function, BatchNorm1d for feature normalization, and AvgPool1d to drastically reduce the latent spaces. The model has a total of 474,562 parameters. As a side note, a Bayesian model requires two times more weights than a deterministic one, because each weight would be represented by a mean ($\mu$) and a standard deviation ($\sigma$), to define its normal distribution. It was optimized using AdamW optimizer, with a cyclical annealing to enhance the ELBO loss optimization, and a learning rate scheduling containing a warm up and then an exponential learning rate scheduler. A batch size of 21, 000 is applied, accounting for approximately 50\% of the largest experiments in the training.

As illustrated in Figure~\ref{fig:training_phase_losses}, the training process is monitored by tracking the value of the different losses. The primary objective is to minimize the Evidence Lower Bound (ELBO), which can be decomposed into two parts with different roles. The correlation value tracking during training for our baseline model is also available in in Figure~\ref{fig:train_corr} .

The first component is the reconstruction loss (or data likelihood term). Its purpose is to quantify how well the model's predictions fit the observed data. Minimizing this term drives the model to learn accurate representations.

The second component is the Kullback–Leibler (KL) divergence term. This acts as a regularizer by measuring the distance between the learned posterior distribution of the weights and a simple prior distribution. Minimizing the KL divergence prevents overfitting by ensuring that the model does not diverge too far from its prior assumptions, thereby promoting generalization.

Often, the learning process only optimizes the KL loss, leading to an useless training. Several techniques, such as annealing or regularization can help solve this problem and enable the reconstruction learning process.

\section{Multinomial model to Poisson model}
\label{sec:mult_to_poisson}

\begin{figure}
    \centering
    \includegraphics[width=0.95\linewidth]{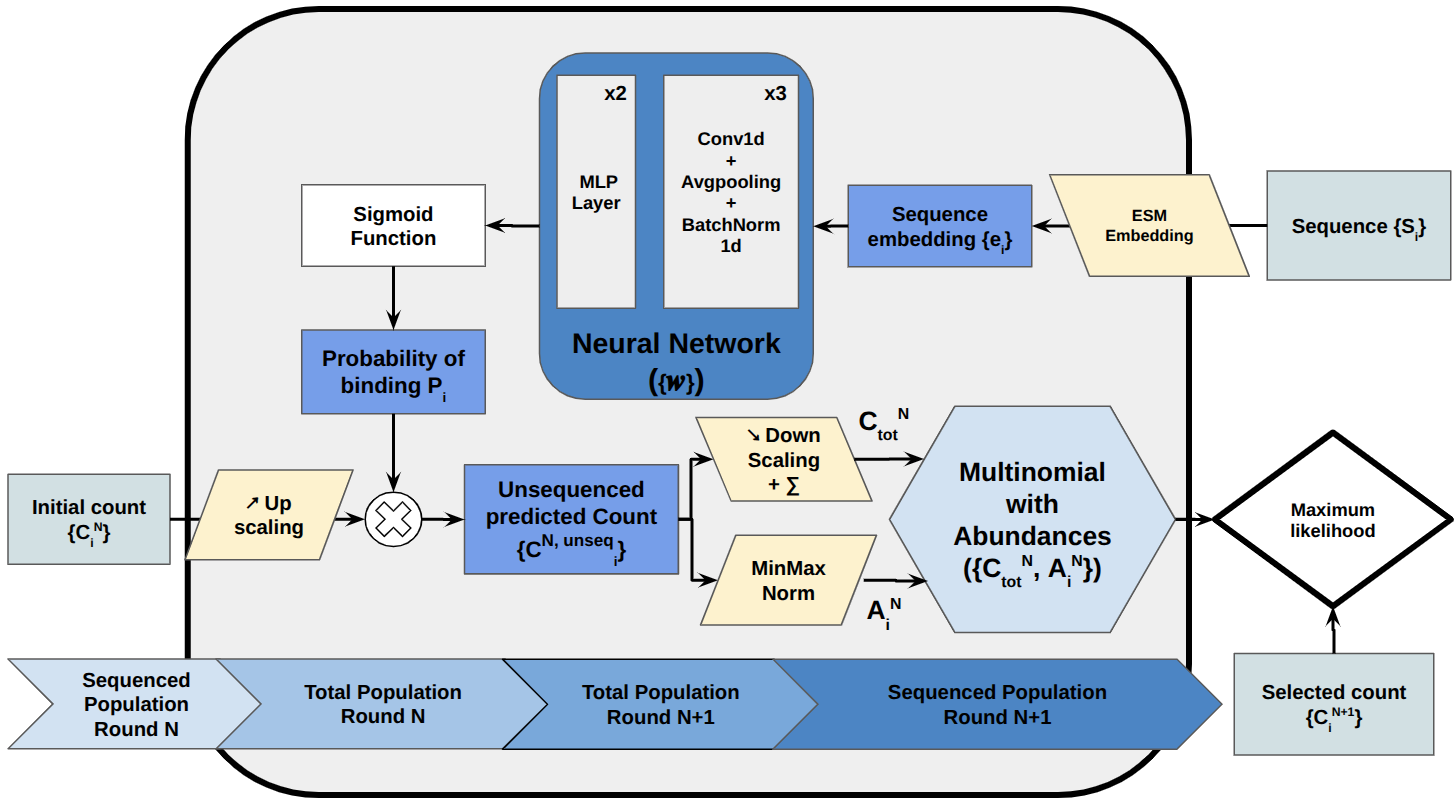}
    \caption{\textbf{Phage display Multinomial model}}
    \label{fig:multinomial}
\end{figure}

The initial approach to modelling sequence subsampling was based on the \textbf{multinomial distribution}, as it naturally accounts for multiple outcomes from a fixed number of trials. This is depicted in Figure~\ref{fig:multinomial}. However, this model's inherent dependencies among variables and its computational complexity led to a significant bottleneck, particularly with large experiments, as this dependency constrains a batch to contain a whole experiment.

A key observation that led to the bypass of this issue was the nature of our total count, $N$. As $N$ is extremely large and imprecise (approximately $10^{13}$), we can approximate its distribution with a \textbf{Poisson distribution}, using the total count itself as the rate parameter, $\lambda$. This approximation is the foundation for a more tractable model.

This strategic choice enables a powerful mathematical transformation known as \textbf{Poissonization}. By modelling the total count as a Poisson random variable, we can exactly transform the dependent multinomial variables into a set of independent Poisson variables. This transition from a dependent to an independent framework dramatically simplifies subsequent calculations and resolves the initial computational bottleneck.

Proof:

Let $Y = (Y_1, \dots, Y_k)$ be a vector of counts where the total count $N = \sum_{i=1}^k Y_i$ follows a Poisson distribution with parameter $\lambda$. Given $N=n$, the counts follow a multinomial distribution:
$$P(Y_1=y_1, \dots, Y_k=y_k \mid N=n) = \frac{n!}{y_1! \dots y_k!} p_1^{y_1} \dots p_k^{y_k}$$
where $\sum p_i = 1$ and $\sum y_i = n$.

The joint unconditional probability is found by combining the multinomial and the Poisson distributions:
\begin{align*}
P(Y_1=y_1, \dots, Y_k=y_k) &= P(Y_1=y_1, \dots, Y_k=y_k \mid N=n) \cdot P(N=n) \\
&= \frac{n!}{y_1! \dots y_k!} p_1^{y_1} \dots p_k^{y_k} \cdot \frac{e^{-\lambda} \lambda^n}{n!} \\
&= \frac{1}{y_1! \dots y_k!} (p_1 \lambda)^{y_1} \dots (p_k \lambda)^{y_k} e^{-\lambda}
\end{align*}
Letting $\lambda_i = p_i \lambda$, and noting $\sum \lambda_i = \lambda$, we can split the exponential term:
$$P(Y_1=y_1, \dots, Y_k=y_k) = \left( \frac{\lambda_1^{y_1} e^{-\lambda_1}}{y_1!} \right) \left( \frac{\lambda_2^{y_2} e^{-\lambda_2}}{y_2!} \right) \dots \left( \frac{\lambda_k^{y_k} e^{-\lambda_k}}{y_k!} \right)$$
This is the product of the probability mass functions of $k$ independent Poisson distributions. Therefore, each $Y_i$ is an independent Poisson random variable with parameter $\lambda_i$.

\begin{figure}
    \centering
    \includegraphics[width=0.75\linewidth]{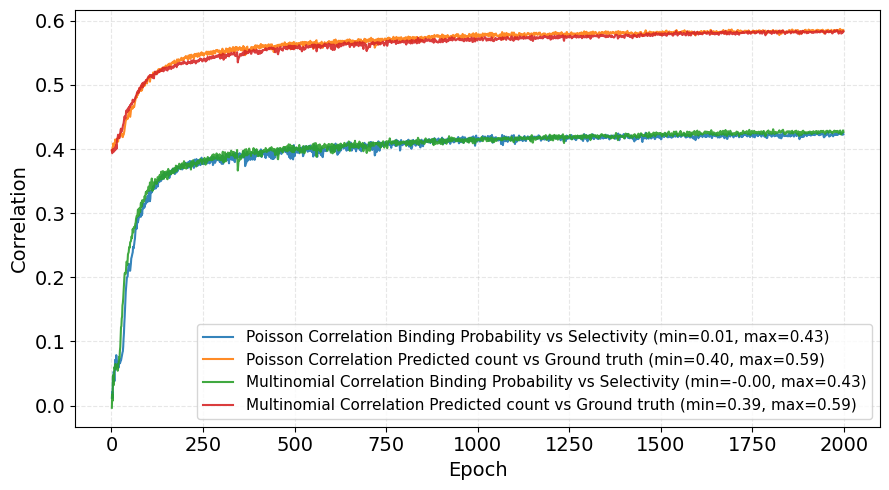}
    \caption{\textbf{Empirical Equivalence of Poisson and Multinomial on Model Performance}.}
    \label{fig:poisson_vs_multinomial}
\end{figure}

Concerning the experimental result, it was observed that the training process and the final results were identical for both the multinomial and Poisson distribution, with the same hyper parameters. This consistency, as shown in Figure~\ref{fig:poisson_vs_multinomial} for one training experiment, provides a strong empirical validation of our approach. The same consistency can be observed for every training and validation correlations.

\section{Batch size study}
\label{sec:batch}


As shown in the upper appendix section, we enabled batch training with Poisson law. But, before the stochastic pass, we encounter a downsampling to the N+1 sequenced population scale, which needs the total sum of the unsequenced survivor population scale. When using stochastic mini batches, this particular total sum is not directly accessible, so, an approximation needs to be done on this total unsequenced count. 

\begin{figure}
    \centering
    \includegraphics[width=1\linewidth]{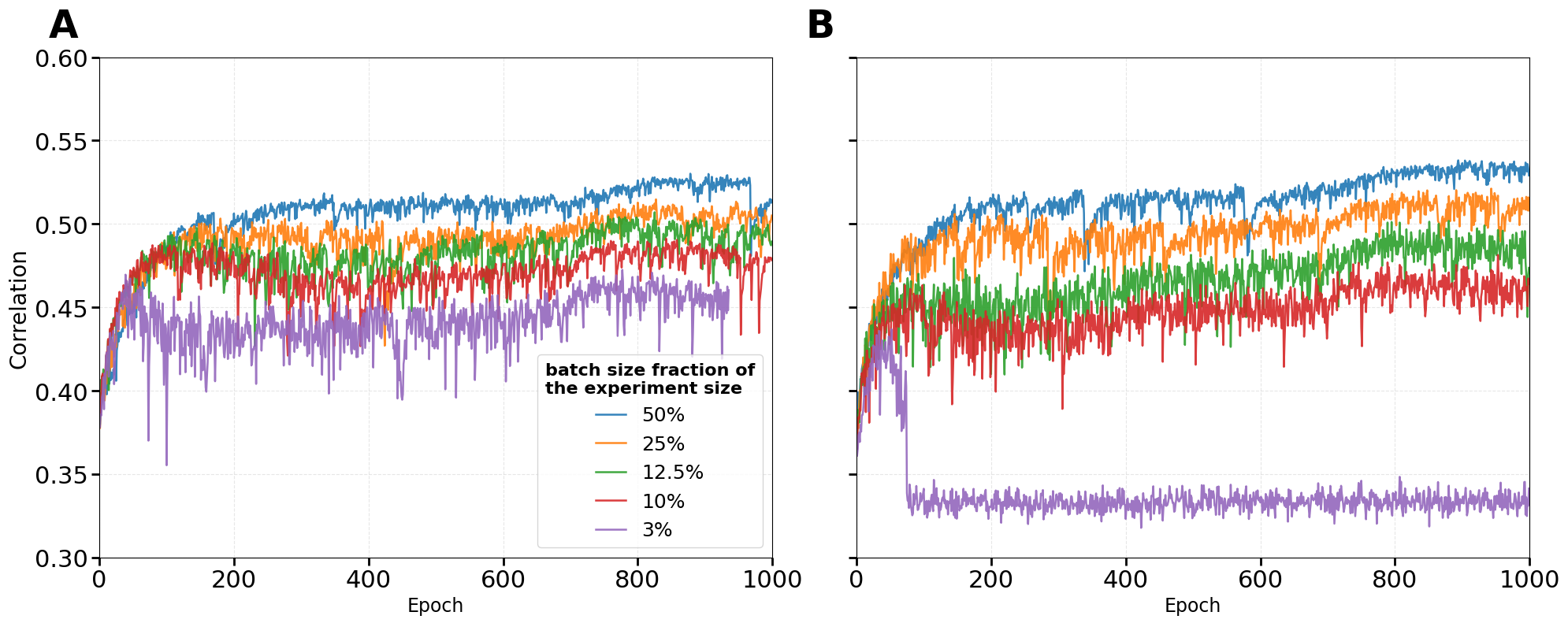}
    \caption{\textbf{Impact of batch size on model training and performance}. \textbf{(A)}: Spearman correlation between Ground truth and Predicted count with Exponential Moving Average approximation method. \textbf{(B)}: Spearman correlation between Ground truth and Predicted count with Naive method.}
    \label{fig:batch_study}
\end{figure}

This approximation gives the batch size hyper-parameter a significant influence on model's performance, which requires further study. The results for various batch sizes are provided in Figure~\ref{fig:batch_study}. We observed that the asymptotic correlation of the model's performance decreases as the batch size becomes a smaller fraction of the total experiment size. Furthermore, a smaller batch size leads to an increased training instability, a known effect that can also act as a form of regularization. In certain scenarios, this regularization effect can be beneficial and enables learning.

Several approximation methods were tested, such as the naive approximation (assuming that the mini batch is representative of the entire population), which would just increase to the total population, for instance, with a randomized batch, if the batch size is $\frac{1}{x}$ of the population, this will translate to multiplying by $x$ to rescale to the total population which, assuming batches being independent, would have approximately the same sum output (see Figure \ref{fig:batch_study} B). A second idea is to take the total sum calculated at the precedent step with, for first value, the naive method applied. Finally, a Moving Average, or an Exponential Moving Average could also be used (see Figure \ref{fig:batch_study} A). Empirical analysis reveals that only the Exponential Moving Average (EMA) offers a notable performance improvement over the naive approach, particularly with smaller batch sizes. The naive method is effective only with large batch sizes, approaching 50\% of the total experimental sample size. However, its performance degrades significantly as the batch size is reduced. Conversely, the superiority of the EMA becomes more pronounced as the batches get smaller. In very low batch size regimes, EMA enables effective learning where the naive method fails entirely. (See Figure~\ref{fig:batch_study})

\section{Multiple output model symmetry and collapse}
\label{section:multiple}

As discussed in the paper, our work addresses two key issues with phage display data. First, the inherent noise is handled effectively by our Bayesian approach. Second, a more fundamental challenge arises from our sequencing protocol, which occurs only once per round. This prevents a clear, experimental separation of negative and positive selection.

To tackle this, we explored various models, as detailed in Section~\ref{Sec:Discussion}, featuring one or more networks designed to output two or four logits. Theoretically, these logits should allow the model to disentangle the two selections using only the mathematical relationships we provided.

However, a significant practical issue arises from the non-uniqueness of the solution. As illustrated by Equation~\ref{eq:pos}, a single observable probability, $P_i$, can correspond to an infinite number of combinations of positive and negative selection probabilities \{$P^{pos}_i$, $P^{neg}_i$\}. This inherent ambiguity means that the model cannot reliably converge on a single, true biological solution.

\begin{figure}
    \centering
    \includegraphics[width=1\linewidth]{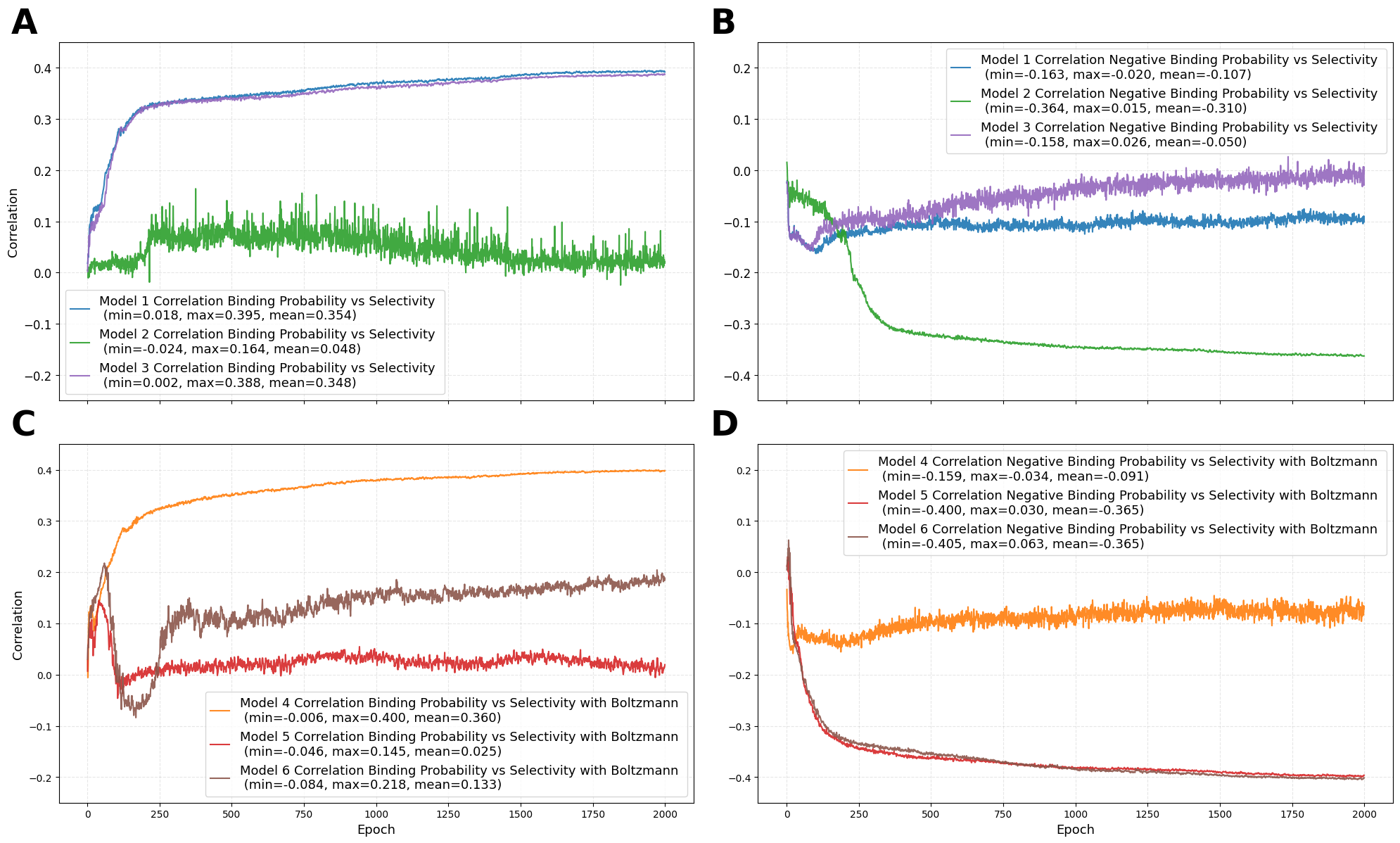}
    \caption{Model Comparison with different Selection Separation and hyper parameters, such as the number of parameters or number of networks. \textbf{(A)}: Positive binding probability for naive models. \textbf{(B)}: Negative binding probability for naive models. \textbf{(C)}: Positive binding probability for Boltzmann models. \textbf{(D)}: Negative binding probability for Boltzmann models.}
    \label{fig:collapsing}
\end{figure}

As shown in Figure~\ref{fig:collapsing}, this issue leads to a wide variety of learning behaviours and final solutions across different model runs. The figure compares three models using the naive approach and three using the Boltzmann model. In most cases, the model "collapses" and relies predominantly on only one of the two selection probabilities. For instance, in Figure~\ref{fig:collapsing}B, model 3 shows an asymptotic correlation of zero, indicating that it exclusively uses what we termed "positive selection probability." In practice, however, we cannot apply this name with certainty, as the model's learned representation may not correspond to the true biological process.

\section{BFN generated sequences}
\label{section:BFN}
We fine tuned the foundational Bayesian Flow Network (BFN) model \citep{Atkinson2025} with a subset of sequences from the lowest target concentration of $\mathrm{experiment_1}$ round 3c. Typically we only considered sequences appearing with a count superior to 40.
Then this fined tuned version of BFN was used for in painting seed sequences from the Phage display data.\\
Seeds were chosen following two paths.
10 were chosen because of their low noise over signal ratio both in our model and in term of selectivities, high expected binding affinity, while spanning a large predicted probability of binding dynamical range, diversity of CDR3. 8 more were chosen from previous wet lab experimental characterization (cloning out), privileging here again spanning the binding affinity dynamical range.\\
In painting was done following 3 different strategies:
\begin{itemize}
    \item only CDR3
    \item all CDRs
    \item all CDRS and framework 3
\end{itemize}

\begin{figure}[h]
    \centering
    \includegraphics[width=0.95\linewidth]{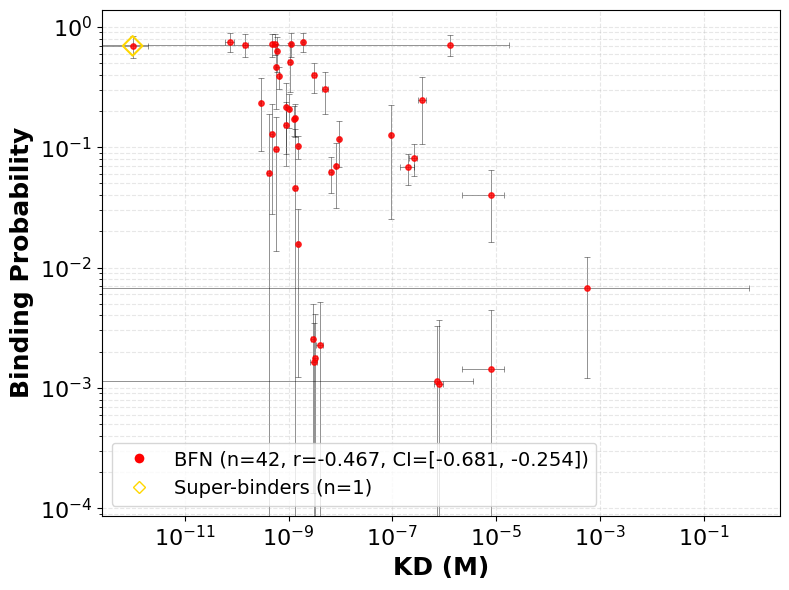}
    \caption{\textbf{Correlation plot for the dissociation constant test set, using only the BFN values}}
    \label{fig:BFN}
\end{figure}

This has lead to roughly 1800 generated sequences from which we subsampled greedily for diversity, predicted binding probability and predicted sequence naturalness, to end up with the sequences that have been presented here. Here shown in the Figure~\ref{fig:BFN}, the result of our model on only the  BFN data points $K_d$, where we can see that our artificial intelligence model seems to perform better on AI generated sequences than on the whole set of experimental measurement comprising also natural sequences.

\section{Experimental setup}
\label{app:experimental}

For Target 1, the training set included Exp. 1, 2, 4, and 5, with reported training metrics corresponding solely to Exp. 1 and validation performed on Exp. 5. Similarly, Target 2 was trained on Exp. 1 and 2 (metrics presented here from Exp. 1) and validated on Exp. 3. Target 3 was trained on Exp. 5, 6, 11, and 12 (metrics presented here from Exp. 5) and validated on Exp. 7. For the Test set, the selectivities that we compare to the $K_d$ values are aggregating together every experiments from Target 1.

\begin{figure}[h]
    \centering
    \includegraphics[width=0.95\linewidth]{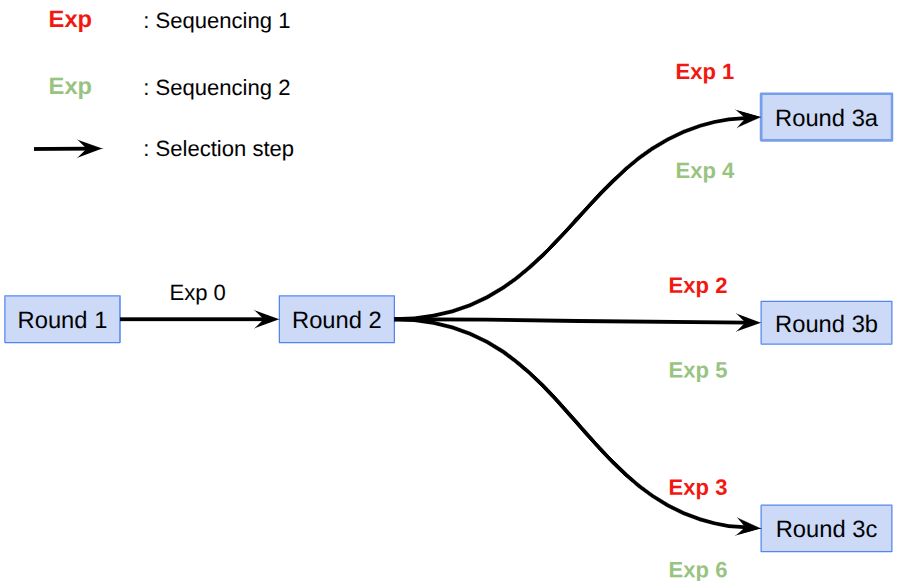}
        \caption{\textbf{Schematic of} $\mathbf{experiment_1}$ ($\mathbf{target_1}$). a,b,c refers to 3 different target concentrations, a being the highest concentration and c the lowest. Rounds 3 are sequenced twice with different PCR protocols.}
    
    \label{fig:experiment1}
\end{figure}

\begin{figure}[h]
    \centering
    \includegraphics[width=0.95\linewidth]{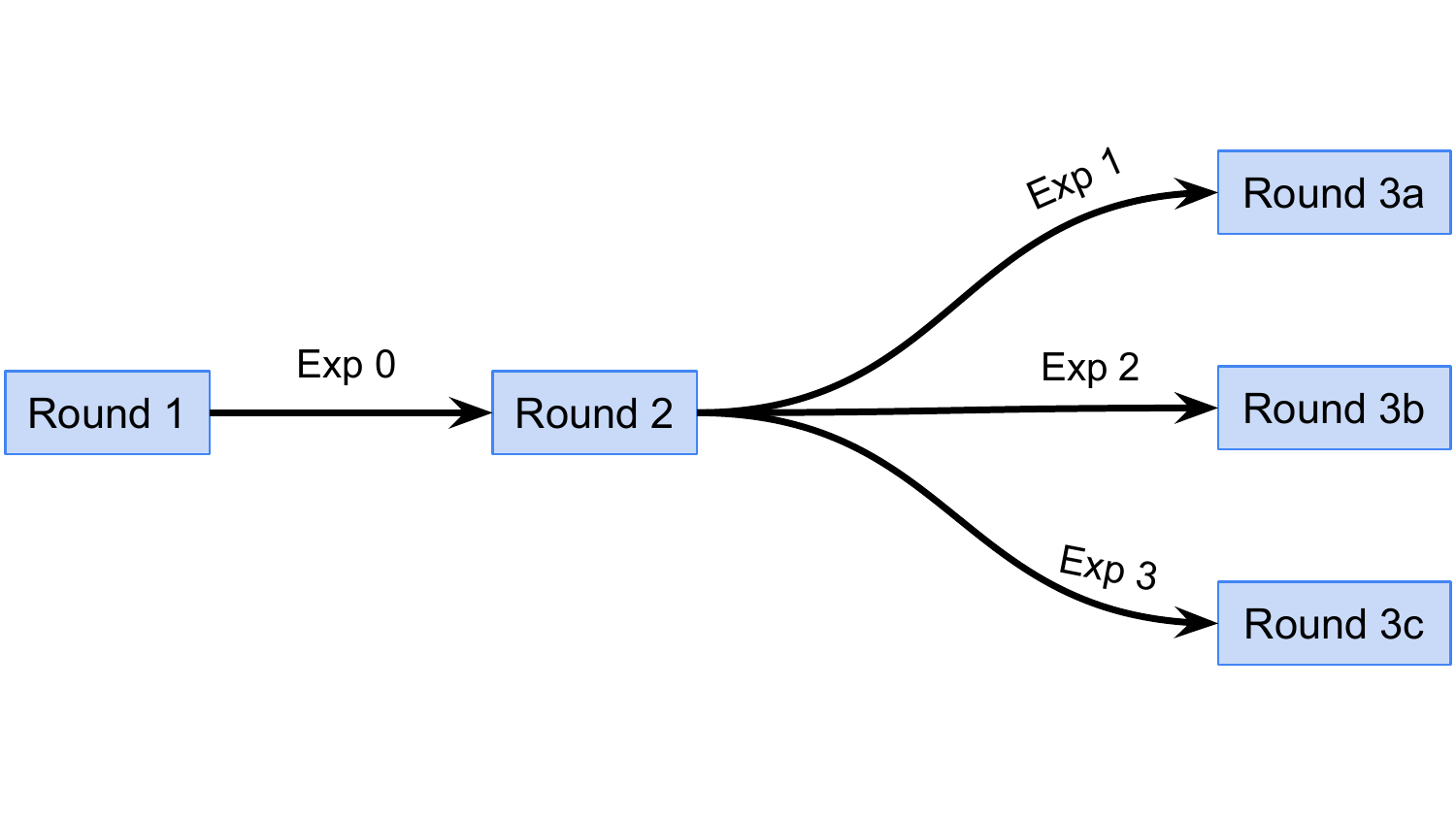}
        \caption{\textbf{Schematic of} $\mathbf{experiment_2}$ ($\mathbf{target_2}$). a,b,c refers to 3 different target concentrations, a being the highest concentration and c the lowest.}
    
    \label{fig:experiment2}
\end{figure}

\begin{figure}[h]
    \centering
    \includegraphics[width=0.95\linewidth]{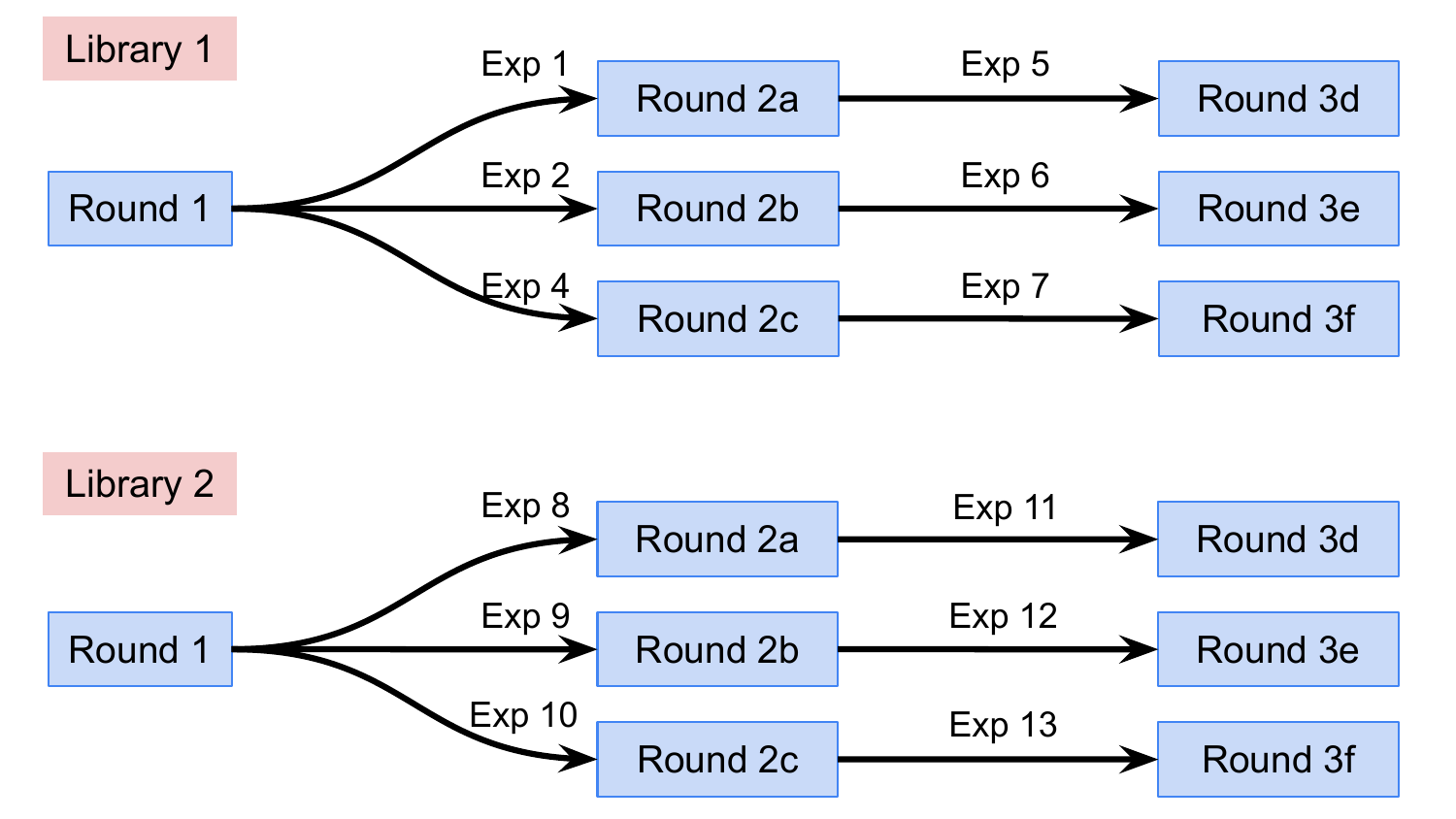}
        \caption{\textbf{Schematic of} $\mathbf{experiment_3}$ ($\mathbf{target_3}$). a,b,c refers to 3 different target concentrations for round 2, a being the highest concentration and c the lowest. d,e,f refers to 3 different target concentrations for round 3, d being the highest concentration and f the lowest. Target concentration at round 3 are always smaller than at their respective round 2. Moreover for this particular target two libraries were derived and tested.}
    
    \label{fig:experiment3}
\end{figure}

\begin{figure}[h]
    \centering
    \includegraphics[width=1\linewidth]{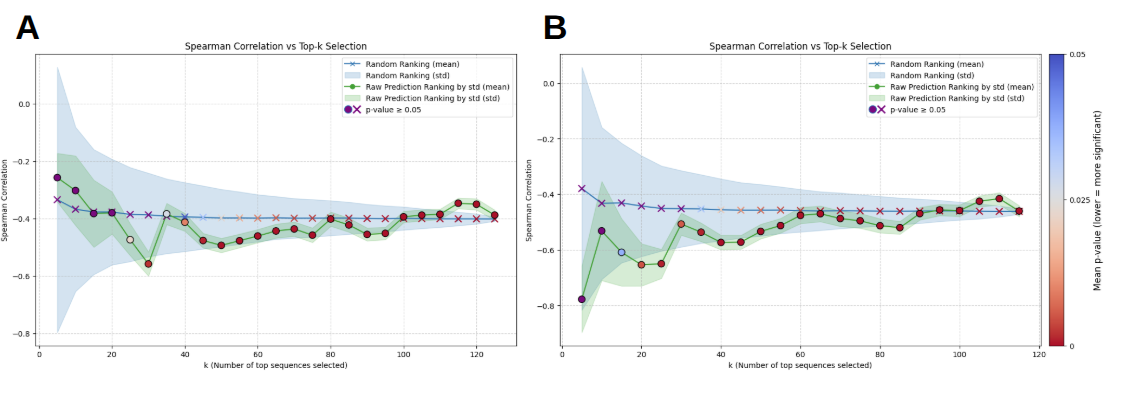}
    \caption{
        \textbf{Model-Guided with Relative Error vs. Random Top-k Enrichment Performance.}
        Comparison of Spearman's rank correlation achieved by two sampling strategies as the sample size is iteratively increased. 
        The dotted line shows the spearman correlation metric between $K_d$ values and model's output on the specific sample where candidates are prioritized by the relative error from our Bayesian Neural Network; its shaded region denotes one standard deviation derived from 1000 samples of the BNN's posterior, representing model uncertainty.
        The crossed line represents the baseline performance, averaged over 1000 independent random sampling runs; its standard deviation reflects the variability inherent to the random sampling process.
        \textbf{(A)} When performed on the full test set (126 compounds), the BNN-guided strategy's performance is initially skewed by a known model failure mode: the misranking of a "super-binder" within the top candidates, which heavily penalizes the correlation score. 
        \textbf{(B)} Upon removal of these outliers, the superiority of the error-based proxy becomes evident, as the BNN-guided enrichment significantly and consistently outperforms the random sampling baseline. 
        This result highlights that the BNN's relative error is a robust metric that can effectively guide experimental decision-making.
    }
    \label{fig:top_k_enrichment}
\end{figure}

\begin{figure}[h]
    \centering
    \includegraphics[width=1\linewidth]{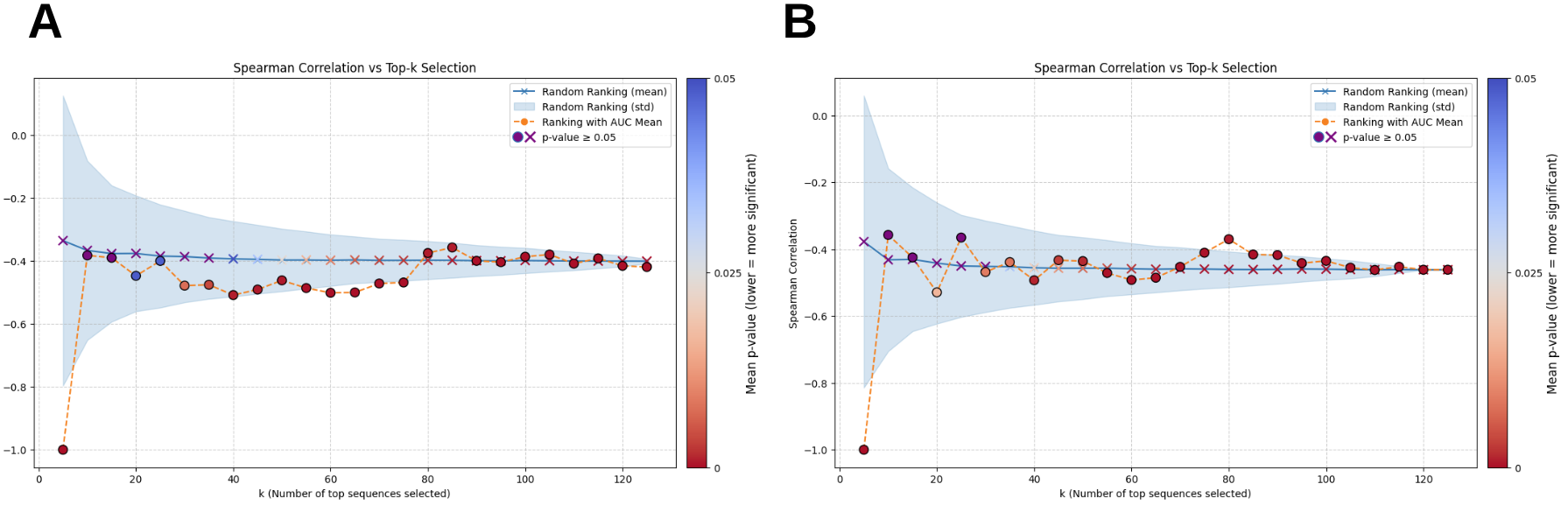}
    \caption{
    \textbf{Model-Guided with AUC Mean vs. Random Top-k Enrichment Performance}
    Same experimental setup as in Figure~\ref{fig:top_k_enrichment}, but using the AUC Mean from Integrated Gradient Explainability method as the guiding metric for candidate selection instead of the relative error. \textbf{(A)}: This result, with the total Test Set, highlights that the AUC Mean metric is a robust value that can effectively guide experimental decision-making. \textbf{(B)}: In contradiction, the selection efficiency collapses to near-random levels across the majority of the curve when super-binders are withdrawn from the Test Set. This observation may suggests that the high initial utility of the AUC Mean metric in Panel (A) was predominantly driven by its ability to correctly identify the extreme case. Once these high-leverage points are removed, the metric fails to effectively discriminate between candidates. This performance drop is counter-intuitive and requires further investigation, as it implies a fundamental limitation in the AUC Mean metric's ability to guide the most relevant output of our model.
}
    \label{fig:top_k_enrichment_auc}
\end{figure}

\begin{figure}[h]
    \centering
    \includegraphics[width=0.95\linewidth]{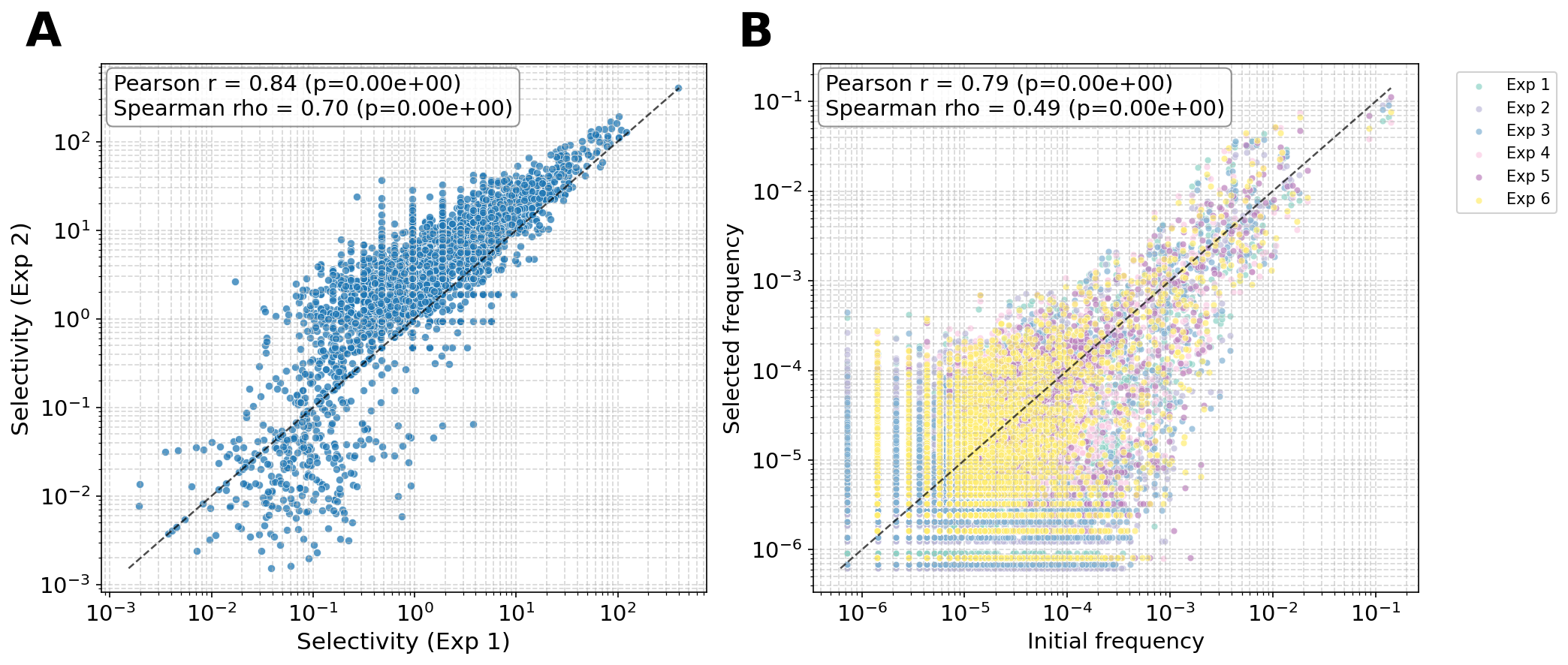}
    \caption{\textbf{Correlations across the different experiments for target 1}. The data shows the experiments are highly correlated. \textbf{(A)}: Scatter plot of the selectivity from Exp. 1 (round 3a) versus Exp. 2 (round 3b). \textbf{(B)}: Initial frequency (round 2) versus selected frequency for each experiment (Exp 1-3 = rounds 3a-c) and Exp 4-6 =  technical sequencing replicates of rounds 3a-3c), showing that the distributions are quite similar. The relationship between selectivity and frequency is described in Equation~\ref{eq:select}.}
    \label{fig:selectivity}
\end{figure}

\begin{figure}[h]
    \centering
    \includegraphics[width=1\linewidth]{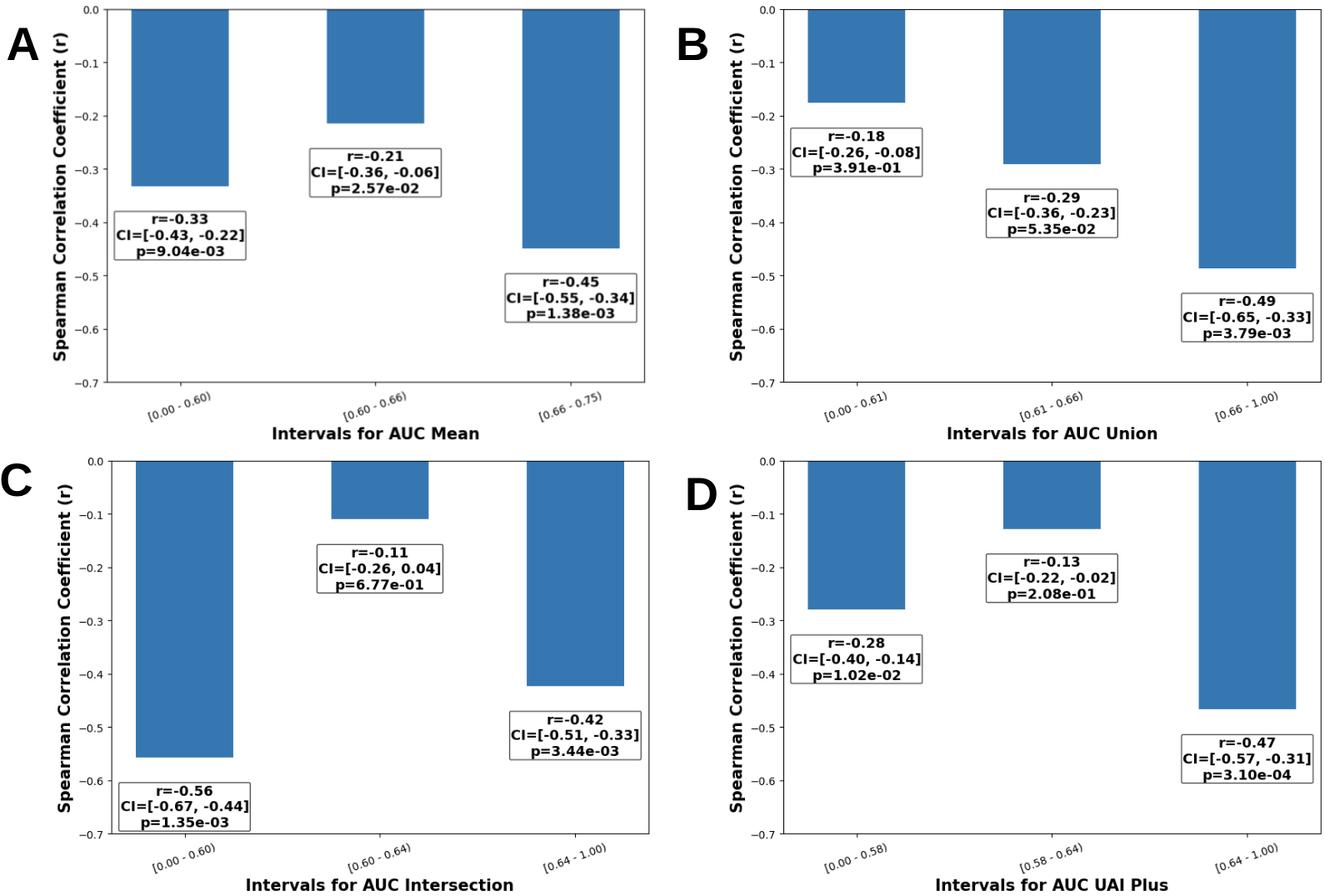}
    \caption{\textbf{Comparison of the different Explainability aggregation on AUC metric}. We evaluate four distinct methods for aggregating feature attributions generated using Integrated Gradients across 100 posterior samples from a Bayesian Neural Network (BNN). The performance, measured by AUC, highlights the unique characteristics of each aggregation strategy. \textbf{(A) Mean Aggregation:} This approach calculates the simple average of attribution scores across all 100 BNN samples, providing a general measure of feature importance. The results do not show a clear monotonic progression through the score bins; however, the bins with higher average scores tend to yield better performance. This suggests a potential threshold effect, where features must achieve a certain level of mean importance to be reliably impactful. \textbf{(B) Union Aggregation:} This method uses a high percentile of the attribution distribution to identify features considered important in at least one of the BNN samples. It effectively captures any feature that is strongly influential, even if inconsistently. This strategy demonstrates a clear and consistent progression in performance, with the AUC improving as the aggregated scores increase, proving its effectiveness in guiding toward reliable predictions. \textbf{(C) Intersection Aggregation:} In contrast, this approach uses a low percentile to identify features that are consistently non-influential or have negligible importance. The results show a counter-intuitive inverse trend, where the lowest-scoring bin achieves the best AUC. This indicates that the metric is ill-suited for identifying positive predictive features, as it is designed to highlight negligible or even negative contributions to the model's output. \textbf{(D) UAI Plus Aggregation:} This method measures the frequency of importance by calculating the percentage of the 100 samples where a feature's absolute attribution value exceeds a predefined threshold, $\tau$ (i.e., \% of samples where $|attribution|>\tau$). This provides an interpretable metric of how consistently a feature is considered important. The choice of hyper-parameter $\tau$ makes this aggregation highly dependent, and will need to be tuned for each new XAI method. The performance of this method is largely comparable to that of the Mean Aggregation, suggesting a similar overall behaviour in this evaluation context.}
    \label{fig:auc_comparison}
\end{figure}

\begin{figure}[h]
   \centering
    \includegraphics[width=0.4\linewidth]{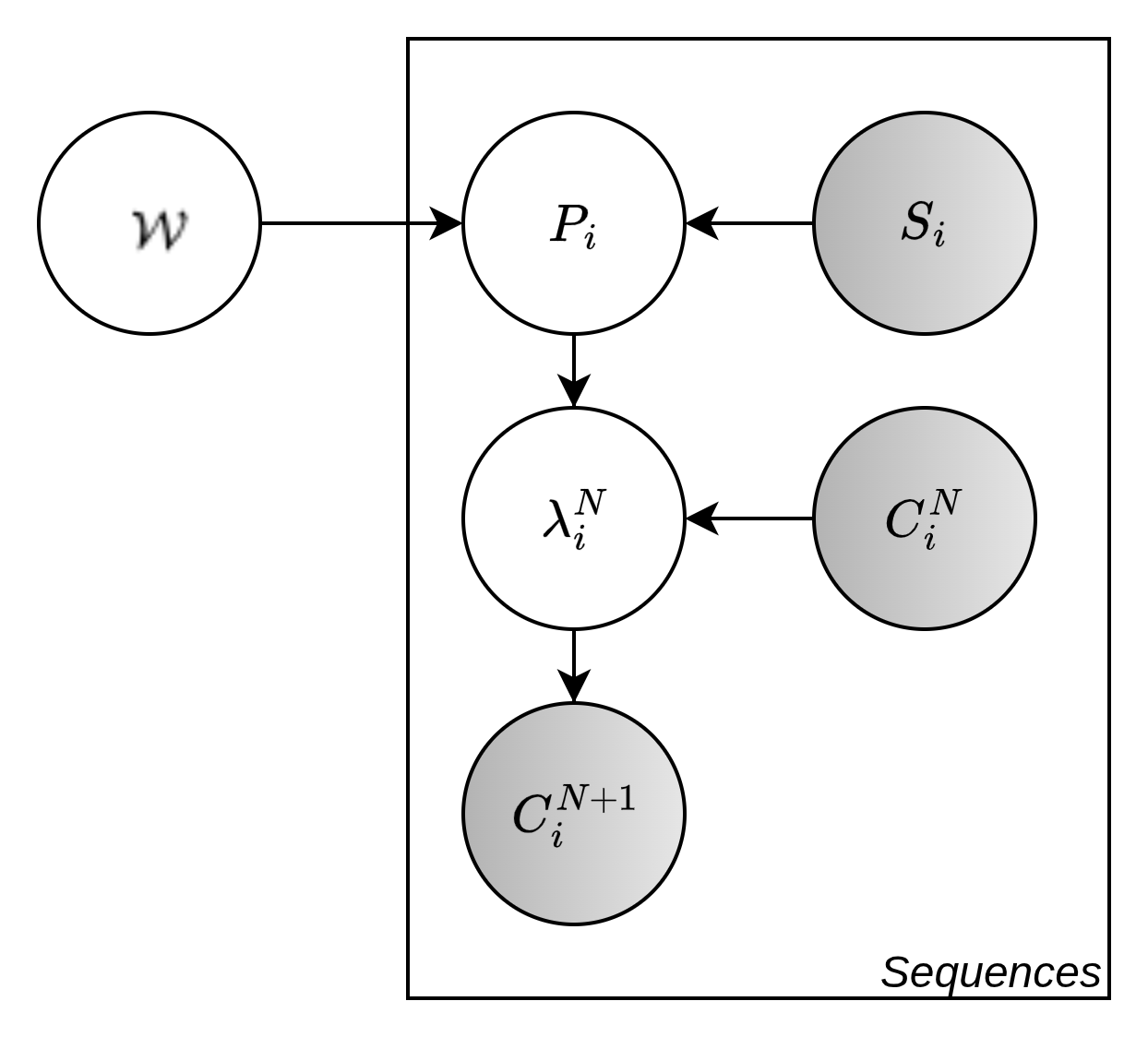}
        \caption{\textbf{Bayesian Network of our Poisson law model}. This diagram illustrates the dependencies between the variables in our Bayesian model. The nodes represent the following: \textbf{$\mathcal{W}$} are the weights of the Bayesian neural network; \textbf{$P_i$} is the probability of binding for sequence $i$; \textbf{$S_i$} denotes the amino acid sequence $i$; \textbf{$\lambda_i^N$} is the rate parameter for sequence $i$ at round $N$ of the corresponding Poisson distribution; \textbf{$C_i^N$} is the count for sequence $i$ at round $N$; and \textbf{$C_i^{N+1}$} is the predicted count for sequence $i$ at the next round, $N+1$. The shaded nodes, $S_i$ and $C_i^N$, are observed, while the unshaded nodes are latent.}

   \label{fig:bayes}
\end{figure}

\begin{figure}[h]
    \centering
    \includegraphics[width=0.95\linewidth]{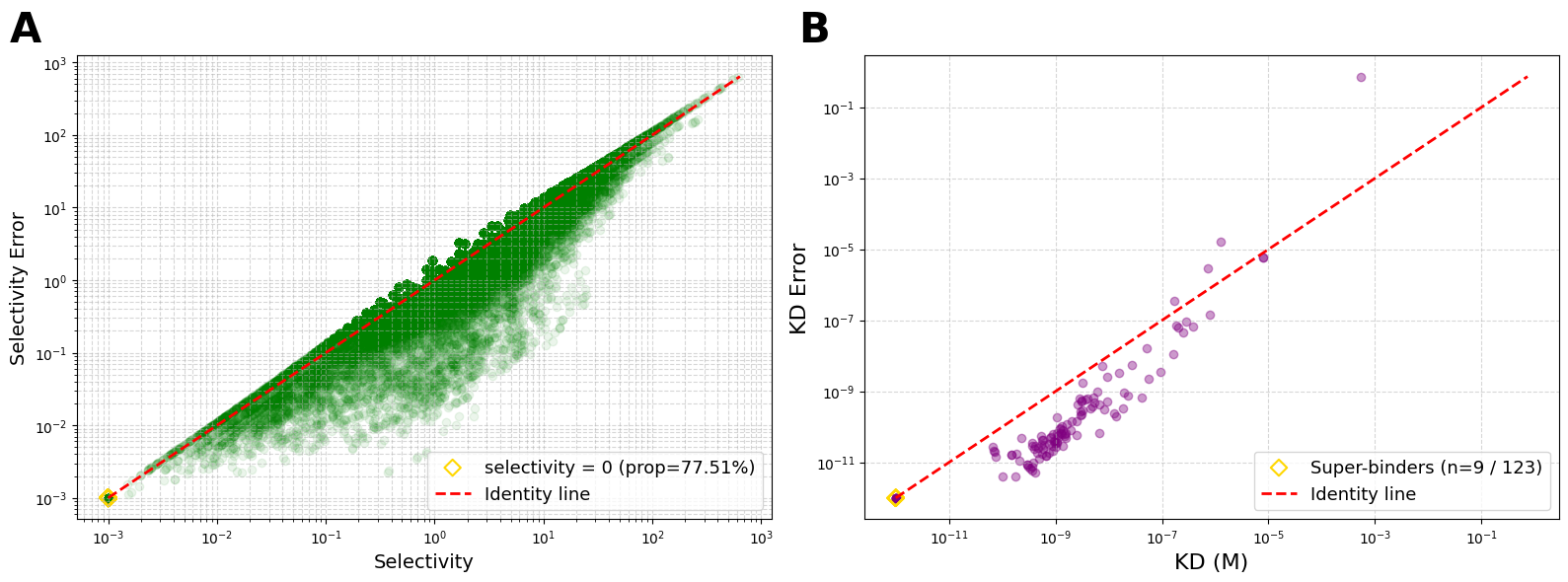}
        \caption{\textbf{Characterization of Measurement Uncertainty for Selectivity and Dissociation Constant.} The provided plots characterize the measurement noise for target 1 by displaying the data versus its corresponding error on a log-log scale. The concentration of points near the identity line reveals a consistently high relative error throughout the dataset. \textbf{(A)}: The error in selectivity $\Delta s$) is estimated assuming Poisson-distributed counts, using the approximation: $\frac{\Delta s}{s} = \frac{1}{\sqrt{C_i^N}} + \frac{1}{\sqrt{C_i^{N+1}}} + \mathcal{O}(C_{tot}^N)$.
        However, this approximation is inadequate for a large portion of the data, as over 77\% of the entries exhibit a selectivity of zero. For these points, the error is undefined, presenting a significant challenge for modelling. \textbf{(B)}: A similar trend of high uncertainty is observed for the dissociation constant ($K_d$). Notably, "super-binders" exhibit particularly large errors. This is attributed to their high affinity, which saturates the instrument's detectors and caps the measurements at the limit of the device's dynamic range.}
    
    \label{fig:noise}
\end{figure}




\end{document}